\documentclass{article}
\newcommand{\eb}{\ee\be } 
 
\newcommand{\bmat}{\lt ( \begin{array} }
\newcommand{\emat}{  \end{array} \rt )}

\newcommand{\oH}{{\ov H}}
\newcommand{\oP}{{\ov P}}

\newcommand{\oQ}{{\ov Q}}

\newcommand{\oR}{{\ov R}}

\newcommand{\oB}{{\ov B}}

\newcommand{\oJ}{{\ov J}}
\newcommand{\cd}{{\cdot}}

\newcommand{\ob}{{\ov b}}

\newcommand{\ovv}{{\ov v}}
\newcommand{\ot}{{\ov t}}

\newcommand{\oT}{{\ov T}}
\newcommand{\oG}{{\ov \G}}
\newcommand{\ovG}{{\ov G}}
\newcommand{\oK}{{\ov K}}

\newcommand{\ED}{\end{document}}

\newcommand{\oY}{{\ov Y}}
\newcommand{\og}{{\ov g}}
\newcommand{\oy}{{\ov \y}}

\newcommand{\oC}{{\ov C}}
\newcommand{\oL}{{\ov L}}

\newcommand{\A}{{\ov A}}

\renewcommand{\a}{\alpha}	
\renewcommand{\b}{\beta}

\renewcommand{\d}{\delta}
\newcommand{\e}{\epsilon}
\newcommand{\ve}{\varepsilon}

\newcommand{\f}{\phi}

\newcommand{\y}{\psi}

\newcommand{\G}{\Gamma}

\renewcommand{\S}{\Sigma}

\newcommand{\Y}{\Psi}

\newcommand{\la}{\label}
\newcommand{\ci}{\cite}

\newcommand{\ds}{\documentstyle}	
\newcommand{\fr}{\frac}

\newcommand{\pa}{\partial}
\newcommand{\ov}{\overline}
\newcommand{\be}{\begin{equation}}
\newcommand{\ee}{\end{equation}}
\newcommand{\ba}{\begin{array}} 
\newcommand{\ea}{\end{array}}
\newcommand{\bea}{\begin{eqnarray}}
\newcommand{\eea}{\end{eqnarray}}
\newcommand{\ra}{\rightarrow}
\newcommand{\Ra}{\Rightarrow}

\newcommand{\lt}{\left}
\newcommand{\rt}{\right}

\newcommand{\ben}{\begin{enumerate}}
\newcommand{\een}{\end{enumerate}}
\newcommand{\bitem}{\begin{itemize}}
\newcommand{\eitem}{\end{itemize}}
\begin{document}
\Large

\setlength{\oddsidemargin}{0in}
\setlength{\textwidth}{6.5in}

\begin{center}
{  BRS Cohomology, Composite Operators and the Supersymmetric Standard Model}

\vspace{.2cm}

{\small John  Dixon\footnote{jadix@telus.net}  
\\ \small Dixon Law Firm, 1020 Canadian Centre, 833 - 4th Ave. S. W., Calgary, Alberta, Canada T2P 3T5 }
\\
%\today
\end{center}

%\tableofcontents%\listoftables

{\abstract 
\small
Supersymmetry might be broken, in the real world, by anomalies that affect composite operators, while leaving the action supersymmetric.   New constraint equations that  govern the composite operators and their anomalies are examined. It is shown that the supersymmetric standard model has special properties that allow simple and physically interesting solutions to the constraint equations.}

\twocolumn

\small
%\Large

If supersymmetry \ci{WZ}  is to be a viable theory of the real world, there must exist a mechanism whereby supersymmetry is broken in a compelling and calculable way. It is remarkable that when gauge symmetry is spontaneously broken, in a supersymmetric theory, the vacuum energy remains zero \ci{oRaif}.  Ideally, the mechanism of supersymmetry breaking should maintain this feature. One possibility, which would keep this energy density zero, is that supersymmetry is broken by well-hidden  supersymmetry anomalies in composite operators. 

The state of research on this topic was summarized in 1991 in \ci{dixprl}.   Because of some recent developments to be presented here, the idea seems more promising now than it did in 1991. The most surprising thing  is that the supersymmetric standard model seems to be very well chosen to admit simple and physically interesting solutions to the full cohomology of the Wess-Zumino model.
This is the new result of this paper. 

In \ci{bigpaper} (see also 
\ci{holes}--\ci{Brandt}),  the full cohomology problem for the Wess Zumino model, including composite operators, was analyzed using spectral sequences.  The full nilpotent ($\d^2=0$) operator is set out in the following table (for the massless case):

\vspace{.2cm}

\begin{tabular}{|c|}
\hline
\multicolumn{1}{|c|}{ Table I-A: Transformations $\d$ with $\d^2 =0$}
\\
\hline
{ $\begin{array}{lll}  
\d A^i&= & 
  \y^{i}_{  \b} {C}^{  \b} 
\\
\d \y_{\a}^i &  =& 
\pa_{ \a \dot \b }  A^{i} {\ov C}^{\dot \b}  
+ 
C_{\a}   
G^i
\\
\d \G_i 
&= &
 - \fr{1}{2} \pa_{ \a \dot \b  }       \pa^{ \a \dot \b  }        {\ov  A}_{i} 
+ g_{ijk} { G}^{jk}
+
\pa_{ \a \dot \b } Y_{i}^{ \a}    {\ov C}^{\dot \b}   
\\
\d Y_{i}^{ \a} 
&=&
-
  \pa^{\a \dot \b  }   
{\ov \y}_{i   \dot \b}
+
2 g_{ijk}  \y^{j \a} A^k    
-
\G_i  
 {C}^{  \a}
\\
\d G^i 
&=&
\pa_{\a \dot \b} \y^{i \a} \oC^{\dot \b}
\\
\end{array}$}
\\
\hline
\end{tabular}

\vspace{.2cm}
 $\G$ and $Y$ are Zinn-Justin \ci{ZJ} sources for the supersymmetry variations of the scalar field $A$ and spinor field $\y$, respectively.   $G$ is the composite auxiliary resulting from the integration of the auxiliary $F$. The index $i$ is a general index so that the action contains all possible Wess-Zumino multiplets and all possible dimensionless interactions between them.
The `auxiliary' $G^i$ is composite. Its form, and the first of its higher versions, are indicated in Table I-B. We will introduce a constant spinor source $\f_{\a}$ below.

\vspace{.2cm}

\begin{tabular}{|c|}
\hline
\multicolumn{1}{|c|}{ Table I-B:
 Composite Auxililaries}
\\
\hline
{ $\begin{array}{lll}  
 G^i 
&=&
- \lt ( 
 {\ov g}^{ijk} {\ov A}_j  {\ov A}_k +
{\ov Y}^{i \dot \b} {\ov C}_{\dot \b } 
\rt )
\\
G_2^{ij}&= & 
     A^{i}   
G^j
 +     A^{j}   
G^i
- \y^{i \a} 
\y^{j}_{ \a}  
\\
\end{array}$}
\\
\hline
\end{tabular}

The results of the cohomology analysis of \ci{bigpaper} are summarized  in Table II-A.  That analysis tells us that for an arbitrary massless theory with superspace potential ${\cal Y}$, one should  consider composite spinor expressions of the form\footnote{More generally $\Y= 
f^{j_1 \cdots j_m}_{i_1 \cdots i_n} 
\f^{\a_1 \cdots \a_n} \y^{i_1}_{\a_1} \cdots \y^{i_n}_{\a_n} \A_{j_1} \cdots \A_{j_m}$} 
 $\Y$ and related composite scalar expressions of the form ${\cal H}$.  Then one constructs an operator $d_5$ and its adjoint $d_5^{\dag}$ from the superspace (Yukawa) potential:

\vspace{.2cm}

\begin{tabular}{|c|}
\hline
\multicolumn{1}{|c|}{ Table II-A: Yukawa and Composite Expressions }
\\
\hline
{ $\begin{array}{lll}  
{\ov {\cal Y}} = \og^{ipq} \A_{i} \A_p \A_q 
\\
\Y= 
f_{i}^{jkl} 
\f \cd \y^{i}  
\A_{j}\A_{k}\A_{l} 
 
\\
{\cal H} = 
s^{pqjkl} \f \cd C
\A_p \A_q \A_j \A_k \A_l
\\
d_5 = \lt \{
C_{\a} \og^{ipq} \A_p \A_q 
\rt \}
\y^{i \dag}_{\a}
\\
d_5^{\dag} = 
\y^{i }_{\a}
\lt \{
C_{\a} \og^{ipq} \A_p \A_q 
\rt \}^{\dag}
= \y^{i }_{\a} g_{ipq} C_{\a}^{\dag}  \A_p^{\dag} \A_q^{\dag} 
\\
\end{array}$}
\\
\hline
\end{tabular}

\vspace{.2cm}

Then we must impose the constraints in Table II-B:

\vspace{.2cm}

\begin{tabular}{|c|}
\hline
\multicolumn{1}{|c|}{ Table II-B: Constraint Equations for $\Y$ and ${\cal H}$}
\\
\hline
{ $\begin{array}{lll}  
d_5 \Y= 0
\Ra \f \cd C f^{jkl}_{i} \og^{ipq} \A_p \A_q \A_j \A_k\A_l
=0 
\\
\Ra f^{(jkl}_{i} \og^{pq)i}=0
\\
d_5^{\dag} {\cal H}= 0
\Ra \f \cd \y^{i}
 s^{pqjkl} g_{ipq}  \A_j \A_k \A_l
=0
\\
\Ra s^{pqjkl} g_{ipq} =0
\\
\end{array}$}
\\
\hline
\end{tabular}

\vspace{.2cm}

For each set of solutions $(g_{ijk},f^{lmn}_{p},s^{qrstu})$ of these equations there exist composite operators in the cohomology space of the theory of the following form:

\vspace{.2cm}

\begin{tabular}{|c|}
\hline
\multicolumn{1}{|c|}{ Table II-C}
\\
\hline
\multicolumn{1}{|c|}{ Composite Operators and their Anomalies}
\\
\hline
{ $\begin{array}{lll}  
\Y \ra {\cal O}^{(0)}_{\f} = 
\int d^4 x \; \f^{\a} 
f_i^{jkl}
\lt \{
\oG^{i} \A_j\A_k\A_l
C_{\a} 
\rt.
\\
\lt.
+
\lt (
\pa_{\a \dot \b} A^i 
+ C_{\a}\oY^{i}_{ \dot \b} \rt )
\A_j \A_k \oy_{l}^{\dot \b}
-
\y^i_{\a} 
\ovG_{3,jkl}
\rt \}
\\
{\cal H} \ra {\cal A}^{(1)}_{\f} = 
\int d^4 x \; \f^{\a} C_{\a}
  s^{ijklm}  
\ovG_{5,ijklm}
\\
\d^2=\d {\cal O}^{0}_{\f}=
\d{\cal A}_{\f}^{1}=0
\end{array}$}
\\
\hline
\end{tabular}
\\

\vspace{.2cm}
Then the usual BRS analysis \ci{brs} tells us that for each such ${\cal O}^{0}_{\f}$ with ghost charge zero, there are potential anomalies with ghost charge one, of the form ${\cal A}_{\f}^{1}$.

Our notation for the Superstandard Model is:

\vspace{.2cm}

\begin{tabular}{|c|c|c
|c|c|c
|c|c|c|}
\hline
\multicolumn{8}
{|c|}{Table III}
\\
\hline
\multicolumn{8}
{|c|}{ \bf Superstandard Model, Left ${\cal L}$
Fields}
\\
\hline
{\rm Field} & Y 
& {\rm SU(3)} 
& {\rm SU(2)} 
& {\rm F} 
& {\rm B} 
& {\rm L} 
& {\rm D} 
\\
\hline
$ L^{pi} $& -1 
& 1 & 2 
& 3
& 0
& 1
& 1
\\
\hline
$ Q^{cpi} $ & $\fr{1}{3}$ 
& 
3 &
2 &
3 &
 $\fr{1}{3}$
& 0
& 1
\\\hline
$J$
& 0 
& 1
& 1
& 1
& 0
& 0
& 1
\\
\hline
\multicolumn{8}
{|c|}{ \bf Superstandard Model, Right 
${\cal R}$
Fields}
\\
\hline
$P^p$ & 2 
& 1
& 1
& 3
& 0
& -1
& 1
\\
\hline
$R^p$ & 0 
& 1
& 1
& 3
& 0
& -1
& 1
\\

\hline
$T_c^p$ & $-\fr{4}{3}$ 
& ${\ov 3}$ &
1 &
3 &
 $-\fr{1}{3}$
& 0
& 1
\\
\hline
$B_c^p$ & $\fr{2}{3}$ 
& ${\ov 3}$ 
&
1 &
3 &
 $- \fr{1}{3}$
& 0
& 1
\\
\hline
$H^i$ 
& -1 
& 1
& 2
& 1
& 0
& 0
& 1
\\
\hline
$K^i$ 
& 1 
& 1
& 2
& 1
& 0
& 0
& 1
\\
\hline
\end{tabular}
\\

Here is the superspace potential (Yukawa) for the Standard Model:

\vspace{.2cm}

\begin{tabular}{|c|c|c
|c|c|c
|c|c|c}
\hline
\multicolumn{8}{|c|}{\bf Table IV: Yukawa for Superstandard Model}
\\
\hline
\multicolumn{8}{|c|}{ ${\cal Y} \approx 
g_{ijk}{\cal L}^i {\cal R}^j {\cal R}^k
+ g'' m^2 J
$}
\\
\hline
\multicolumn{8}{|c|}{$
{\cal Y} =
g' \e_{ij} H^i K^j J
+ g'' m^2 J
$}
\\
\multicolumn{8}{|c|}{$
+
p_{pq} \e_{ij} L^{p i} H^j P^q
+
r_{pq} \e_{ij} L^{p i} K^j R^q
$}\\
\multicolumn{8}{|c|}{$
+
t_{pq} \e_{ij} Q^{c p i} K^j T_c^q
+
b_{pq} \e_{ij} Q^{c p i} H^j B_c^q
$}
\\
\hline
\end{tabular}
\\

Note that each term in the Yukawa  for the standard model has the form ${\cal L} {\cal R} {\cal R}$, except for the term  $m^2 J$:  
Hence, for the massless standard model, we note that $d_5$ in Table II-A has the general form
${\ov {\cal R}}^2 C \y_{\cal L}^{\dag}
+
{\ov {\cal L}}{\ov {\cal R}} C \y_{\cal R}^{\dag}$.

Now we shall solve the equations in Table II-B for some examples of $\Y$ and ${\cal H}$ in the {\bf massless} standard model. These solutions will tell us where we can expect to find cohomologically significant composite spinors,  
and also whether those composite spinors can develop anomalies that break their supersymmetry.
The pleasant surprise is that the simplest solutions are composite versions of observable particles. (In fact, the cohomology looks complex enough to encompass all observable particles). In Tables V-${\cal L}$ and  V-${\cal R}$, we write down some examples for possible $\Y$:

{\flushleft
\begin{tabular}{|c|c|}
\hline
\multicolumn{2}{|c|}{ Table V-${\cal L}$: 
 $\Y$ = $\y_{\cal L} {\ov {\cal R}}^{n}$}
\\
\hline
{\rm Name} 
& Composite Spinor 

\\
\hline
Neutrino
&
$
f_{L1,p}
 \f \cd 
\y_{L}^{i  p} 
\oH_i
+
f_{L2}^{p}
 \f \cd 
\y_{J}^{  } 
\oR_p
 $ 
\\
\hline
Proton  
&$
\ve_{c_1 c_2 c_3} 
\lt \{
 \f \cd 
\y_{Q}^{i  c_3 p_3} 
\lt [
f^{p_1 p_2}_{1,p_3}
\oK_i
\oT^{c_1}_{p_1} 
\oT^{c_2}_{p_2} 
\rt.\rt.$\\

\multicolumn{2}{|c|}{ $
\lt.\lt.
+
f^{p_1 p_2}_{2,p_3}
\oH_i
\oT^{c_1}_{p_1} 
\oB^{c_2}_{p_2} 
\rt ]
+
 \f \cd 
\y_{J}
f_3^{p_1 p_2 p_3}
\oB^{c_1}_{p_1} 
\oT^{c_2}_{p_2} 
\oT^{c_3}_{p_3} 
\rt \}
$}
\\
\hline
$\S^{++} $  
&$
\ve_{c_1 c_2 c_3} 
\lt \{
f^{p_1 p_2}_{4 p_3}
 \f \cd 
\y_{Q}^{i c_3 p_3} 
\oH_i
\oT^{c_1}_{p_1} 
\oT^{c_2}_{p_2} 
\rt.
$
\\
&
$
\lt.
+
f_5 \f \cd 
\y_{J}
\ve^{p_1 p_2 p_3}
\oT^{c_1}_{p_1} 
\oT^{c_2}_{p_2} 
\oT^{c_3}_{p_3} 
\rt \}
$
\\
\hline

\end{tabular}
\\
}

\vspace{.01cm}
{\flushleft
\begin{tabular}{|c|c|}
\hline
\multicolumn{2}{|c|}{ Table V-${\cal R}$: 
 $\Y$ = $\y_{\cal R} {\ov {\cal L}}^{n}$}
\\
\hline
{\rm Name} 
& Composite Spinor 

\\
\hline
AntiNeutrino 
&$
f_{L3,p}
 \f \cd 
\y_{R}^{i  p} 
\oJ
+
f_{L4}^{p}
 \f \cd 
\y_{H}^{i  } 
\oL_{ip}
 $ 
\\
\hline
AntiProton &  
$
 \ve^{c_1 c_2 c_3} 
\e^{jk}\oQ_{jc_2 p_2}\oQ_{kc_3 p_3}
$
\\
\multicolumn{2}{|c|}{ $\lt \{
f^{p_1 p_2 p_3}_{6} \f \cd \y_H^i \oQ_{ic_1 p_1} +
f^{p_2 p_3}_{7 p_1} \f \cd \y_{T c_1}^{p_1} \oJ_{} 
\rt \}$}
\\
\hline
\end{tabular}}
\\

\vspace{.2cm}

The constraint from $d_5$ is summarized in Table VI-${\cal L}$ and VI-${\cal R}$ for the two cases.  The identity $\oH_i\oH_j\ve^{ij}=0$ simplifies the result:

\vspace{.2cm}

{\flushleft
\begin{tabular}{|c|c|}
\hline
\multicolumn{2}{|c|}{Table VI-${\cal L}$: 
 $d_5 \y_{\cal L} {\ov {\cal R}}^n =  {\ov {\cal R}}^{n+2}$ }
\\
\hline
{\rm Name} 
& Constraint Equation: 
\\
\hline
Neutrino &  
$
\oH \cd \oK \lt \{
f_{L1,p}
{\ov r}^{pq}
+
f_{L2}^{q}
\og'
\rt \}
\oR_{q}
=0
 $ 
\\
\hline
Proton &  
$
\oH \cd \oK  \ve_{c_1 c_2 c_3} 
\lt \{
\ob^{pq}  \oB^{c_3}_q
f^{p_1 p_2}_{1,p}
\oT^{c_1}_{p_1} 
\oT^{c_2}_{p_2} 
\rt.
$
\\
\multicolumn{2}{|c|}{$\lt.
-
\ot^{pq}   \oT^{c_3}_q
f^{p_1 p_2}_{2,p}
\oT^{c_1}_{p_1} 
\oB^{c_2}_{p_2}+
 \og' f_3^{p_1 p_2 p_3}
\oB^{c_1}_{p_1} 
\oT^{c_2}_{p_2} 
\oT^{c_3}_{p_3} 
\rt \}=0 $}
\\
\hline
$\S^{++} $   &  
$
\oH \cd \oK  \ve_{c_1 c_2 c_3} 
\lt \{
f^{p_1 p_2}_{4 p_3}
\ot^{p_3q}  \oT^{c_3}_q
\oT^{c_1}_{p_1} 
\oT^{c_2}_{p_2} 
\rt.
$ 
\\
&$+
\lt.
\og' f_5
\ve^{p_1 p_2 p_3}
\oT^{c_1}_{p_1} 
\oT^{c_2}_{p_2} 
\oT^{c_3}_{p_3} 
\rt \}=0
$
\\
\hline

\end{tabular}
\\

\vspace{.01cm}

For the VI-${\cal L}$ case, there are plenty of solutions, which could be made more explicit with some straightforward tensor algebra.  Now consider the VI-${\cal R}$ case:

\vspace{.01cm}
{\flushleft
\begin{tabular}{|c|c|}
\hline
\multicolumn{2}{|c|}{Table VI-${\cal R}$: 
 $d_5 \y_{\cal R} {\ov {\cal L}}^n =  {\ov {\cal L}}^{n+1} {\ov {\cal R}}$ }
\\
\hline
{\rm Name} 
& Constraint Equation 
\\
\hline
AntiNeutrino &  
$
f_{L3,p}
{\ov r}^{pq}
\oK \cd \oL_{ q} 
\oJ
+
f_{L4}^{p}
\lt (
\og'
\oK^i \oJ 
\rt.
$
\\
&$\lt.
+
{\ov b}^{rs}
\e^{ij}
\oQ_{jcr} \oB_{s}^{c} 
+
{\ov p}^{rs}
\e^{ij}
\oL_{jr} \oP_{s} 
\rt)
\oL_{ip}
=0
 $ 
\\
\hline
AntiProton &  
$
 \ve^{c_1 c_2 c_3} 
\ve^{im} \oQ_{ i c_1 p_1} \e^{jk} \oQ_{jc_2 p_2}\oQ_{kc_3 p_3} 
$
\\
&$\lt \{
f^{p_1 p_2 p_3}_{6} 
\lt (
\og'
\oK_m \oJ 
+
{\ov b}^{pq}
\oQ_{mcp} \oB_{q}^{c} 
\rt.
\rt. +
$ 
\\
&$
\lt.\lt.
{\ov p}^{rs}
\oL_{mr} \oP_{s} 
\rt)
+
f^{p_2 p_3}_{7 q} \ot^{q p_1} \oK_m   \oJ_{} 
\rt \}=0
$
\\
\hline
\end{tabular}}
\\
\vspace{.1cm}

For the VI-${\cal R}$ case, the equations are too restrictive to admit non-trivial solutions. Next we consider the constraint from $d_5^{\dag}$.  It is summarized in Tables VII-${\cal L}$ and VII-${\cal R}$ for the two cases.
For the ${\cal L}$ case, we see that there are only solutions proportional to $\oJ \oJ$: 

{\flushleft
\begin{tabular}{|c|c|}
\hline
\multicolumn{2}{|c|}{ Table VII-${\cal L}$: ${\cal H}$ = ${\ov {\cal R}}^{n+2} + {\ov {\cal R}}^{n}{\ov {\cal L}}^{2}$ and the effect of $d_5^{\dag}$}
\\
\hline
{\rm Name} 
& Candidate ${\cal H}$ and Solution of $d_5^{\dag} {\cal H}=0$  
\\
\hline
Neutrino &  
$
\oH \cd \oK \oR_{q} s_1^q
+
\oJ \oJ  \oR_{q} s_2^q ;
$\\&$
 s_1^q=0;
 s_2^q$  Free
\\
\hline
Proton &  
$
 \ve_{c_1 c_2 c_3}  
\oT^{c_1}_{p_1}
\oT^{c_2}_{p_2} 
\oB^{c_3}_{p_3} 
 $
\\
 &  
$
\lt \{
h_1^{p_1 p_2 p_3} 
\oH \cd \oK 
+
h_2^{p_1 p_2 p_3} 
\oJ \oJ 
\rt \};
 $\\&$
h_1^{p_1 p_2 p_3}=0; 
h_2^{p_1 p_2 p_3}$ Free
\\
\hline
$\S^{++}$ &  
$
 \ve_{c_1 c_2 c_3}  
\ve^{p_1 p_2 p_3} 
\oT^{c_1}_{p_1}
\oT^{c_2}_{p_2} 
\oT^{c_3}_{p_3} 
 $\\
&$
\lt \{
h_3
\oH \cd \oK 
+
h_4
\oJ \oJ 
\rt \}; 
h_3=0; 
h_4$ Free
\\
\hline
\end{tabular}}

\vspace{.1cm}

On the other hand, for the ${\cal R}$ case in Table VII-${\cal R}$,  there are plenty of solutions, since the equations are not very restrictive:  

\vspace{.01cm}
{\flushleft
\begin{tabular}{|c|c|}
\hline
\multicolumn{2}{|c|}{ Table VII-${\cal R}$}
\\
\hline
\multicolumn{2}{|c|}{Table VII-${\cal R}$: ${\cal H}$ = ${\ov {\cal L}}^{n+1}{\ov {\cal R}}$ and the effect of $d_5^{\dag}$}

\\
\hline
{\rm Name} 
&   Candidate ${\cal H}$ and Solution of $d_5^{\dag} {\cal H}=0$
\\
\hline
AntiNeutrino &  
$
\lt \{
s_3^q
\oK \cd \oL_{ q} 
\oJ
+
s_4^{pqr}
\oQ_{cp} \cd \oL_{q}
\oB_{r}^{c} 
\rt.
$\\&
$
\lt.
+
s_5^{pqr}
\oL_{p} \cd \oL_{q}
\oP_{r} 
\rt \}
 $ 
\\
&
$s_3^r g'
+
 s_4^{prq} b_{pq}
+
 s_5^{prq} p_{pq}=0
;
s_3^r r_{rq}=0 
$\\
\hline
AntiProton &  
$
\ve^{c_1 c_2 c_3} \ve^{ij}  
\oQ_{j c_1 p_1} \oQ_{c_2 p_2} \cd \oQ_{c_3 p_3} 
$\\&$
\lt \{
h^{p_1 p_2 p_3}_{5}  \oJ_{} \oK_i
\rt.
+
h^{p_1 p_2 p_3 p_4 p_5}_{6} \oR_{p_5} \oL_{i p_4}  
$

\\
&$
\lt.
+
h^{p_1 p_2 p_3 p_4 p_5}_{7}   \oQ_{i e p_4 } \oB_{p_5}^e 
\rt \}
$
\\&
$h^{p_1 p_2 p_3}_{5} g'
+
h^{p_1 p_2 p_3 p_4 p_5}_{6} 
r_{p_4 p_5} 
$\\&$+
b_{p_4 p_5}(3 h^{p_1 p_2 p_3 p_4 p_5}_{7} 
+  2h^{p_1 p_2 p_4 p_3 p_5}_{7}
$\\&$
+ h^{p_4 p_2 p_3 p_1 p_5}_{7} ) =0
;h^{p_1 p_2 p_3}_{5} t_{p_1 q}=0
$
\\
\hline
\end{tabular}}
\\
\vspace{.1cm}

In summary then, there are solutions of the constraint equations for composite fermions $\Y$ and for potential anomalies ${\cal H}$.  However, where there are composite fermions in profusion (the $\Y \approx \y_{\cal L}$ case), there is a lack of potential anomalies to be generated\footnote{It looks unlikely  that the $\oJ \oJ$ type anomalies will appear in perturbation theory, since they violate the symmetry of the Yukawa.}. 
And where there is a lack of composite fermions (the $\Y \approx \y_{\cal R}$ case), there are plenty of potential anomalies to be generated, except that there are no operators available  to generate the anomalies.
It follows that, although we have understood the $d_5$ equations better by looking at the standard supersymmetric model, we do not expect to find any supersymmetry anomalies from this insight.
At first sight, this looks like a disappointing result for the attempt to find supersymmetry breaking by this method.

However, this analysis has been done for the massless and ungauged  case only.   When the mass is taken to be non-zero, the vacuum expectation values arise, and spontaneous breaking of SU(2) $\times$ U(1) takes place.  
Once we replace the mass, and then shift the fields $\oK_2 
 \ra m \ovv + \oK_2 ,\oH_1 
 \ra m \ovv + \oH_1 $, to spontaneously break the gauge symmetry, the expressions for the composite fermions ${\cal O}_{\f}^0$ above in Table II-C will typically generate a term in $m$ times the named physical particle.  For example, 
for the neutrino one would expect:
\be
{\rm Neutrino}^{p}
\approx
m {\ov v} 
 \f \cd 
\y_{N}^{p} 
+ O\lt ({\rm Two \;field \;terms}\rt )
\ee

And for the case of $\S^{++} $ , one would expect:
\be m \ovv
\int d^4 x \;
\ve_{c_1 c_2 c_3 }
\f \cd 
 \y_{\rm U,up}^{c_1 }
\oy_{{\rm T,up}}^{c_2} \cd  
\oy_{{\rm T,charm}}^{ c_3}
\eb
+ O\lt ({\rm Four \;field \;terms}\rt ) \ee
This operator creates  the spin one half charmed baryon:
\be
\S_c^{++}(2455) \approx (uuc)_{J^P = \fr{1}{2}^+} 
\ee
and it is one of six possibilities: 
\be
(uuc),
(ucc),
(uut),
(utt),
(tcc),
(ttc)
\ee

It follows from the analysis in Chapter 4 of \ci{bigpaper} that the rest of the supermultiplet that contains this 
$\S_c^{++}(2455)$ can be constructed by letting the source $\f$ become spacetime dependent.  One gets an antichiral spinor superfield multiplet (See \ci{superspace} for some discussion) composed of  a vector, two spinors and a scalar.
If the mechanism of supersymmetry breaking by anomalies works, then the vector and the scalar need to be raised in mass by mixing with the anomaly, leaving the spinor field observable at low energies. 

\vspace{.1cm}

For the ghost charge one sector for this ${\S}^{++}$ case, one needs to extend Table VII-${\cal L}$ to the following:
\[
{\cal H}_{\S}^{++}
=
\f \cd C
\ve_{c_1 c_2 c_3} 
\ve^{p_1 p_2 p_3}
\lt \{
\oT^{c_1}_{p_1} 
\oT^{c_2}_{p_2} 
\oT^{c_3}_{p_3} 
\rt \}\]
\be
\lt \{
h_{3} \ve^{ij}
\oH_i \oK_j
+h_4 \oJ^2 
+
h_5 m^2
+
h_6 m \oJ
\rt \}
\la{massstuff}
\ee
There are also non-gauge invariant terms to add here.  The necessary complete analysis requires a major revision of the whole problem, including gauge fields and spontaneous gauge symmetry breaking etc.  
This changes things so much that the answer is not easy to derive without a lot of careful work which has not yet been done.  
\vspace{.1cm}

Note that one needs three families in order that the expression: 
\be
\ve_{c_1 c_2 c_3} 
\ve^{p_1 p_2 p_3}
\oT^{c_1}_{p_1} 
\oT^{c_2}_{p_2} 
\oT^{c_3}_{p_3} 
\neq 0
\la{threeflavours}
\ee
not vanish.  Hence it appears that there is no possibility of supersymmetry breaking for this multiplet unless there are at least three families.  It is remarkable that the cohomology  depends crucially on  the left--right asymmetry of the theory, and on the direct product nature of the group SU(3) $\times$ SU(2) $\times$ U(1), which results in multiple indices that render the $d_5$ and $d_5^{\dag}$ equations simple to satisfy.  It also appears to be advisable to add the right handed antineutrino multiplet $R^q$ (e.g. for the massive case of Table VII-${\cal L}$).  

\vspace{.1cm}

Assuming that the addition of mass generates a term of the general form of  the potential  anomaly found in equation 
(\ref{massstuff}), we note that (\ref{threeflavours})  is a scalar made with
one top quark or squark (mass 175 GEV), one charm quark or squark,
and one up quark or squark.  So that gives us a first notion of the  magnitude of supersymmetry breaking--the supersymmetry anomaly mixes in at least one top squark or quark with its large mass of 180 GEV, and supersymmetry breaking vanishes as the mass vanishes.
Assuming that the mechanism works, one would expect that the  cosmological constant remains zero after supersymmetry breaking, since the action itself remains supersymmetric.  The supersymmetry breaking would occur only for the composite operators, which do not couple to gravity, and which do not affect the vacuum energy.


\begin{thebibliography}{99}
\bibitem{WZ}J. Wess and B. Zumino, Nucl. Phys. B70(1974) 39.
\bibitem{oRaif} L. O' Raiffeartaigh, Nucl. Phys. B96 (1975) 331.
\bibitem{brs} C. Becchi, A. Rouet, and R. Stora, Commun. Math. Phys. 42 (1975) 127.
\bibitem{superspace} S. J. Gates, M. T. Grisaru, M. Rocek and W. Siegel, Superspace, Benjamin, 1983.
\bibitem{ZJ} J. Zinn-Justin, Quantum Field Theory and Critical Phenomena, Oxford Science Publications, Reprinted 1990.
\bibitem{holes} J. A. Dixon,  Class. Quant. Grav. 7(1990) 1511.
\bibitem{dixprl}  Ibid., Phys Rev. Lett. 67 797 (1991)
\bibitem{kyoto} Ibid. in M. Abe, N. Nakanishi and L. Ojima, BRS Symmetry, Kyoto, Japan (1995) Universal Academy Press Inc., Tokyo
at 143.
\bibitem{bigpaper} Ibid. `Composite Operators, Supersymmetry Anomalies and Supersymmetry Breaking in the Wess-Zumino Model' arXiv-hep-th/0303145 (2003).
\bibitem{dixmin}  J. A. Dixon and R. Minasian, Commun. Math. Phys. 172, 1-11 (1995)
\bibitem{dixminram}  J. A. Dixon, R.  Minasian  and J. Rahmfeld, Commun. Math. Phys. 171, 459-473 (1995)
\bibitem{Rupp} Christian Rupp, Rainer Scharf and Klaus Sibold, Nucl. Phys. B 612 [FS] (2001) 313.
\bibitem{Brandt}F. Brandt,  JHEP04(2003) 035. 
\end{thebibliography}
\end{document}